# Pulsar timing array source ensembles

Bruce Allen[*] and Serena Valtolina[†]

*Max Planck Institute for Gravitational Physics (Albert Einstein Institute), Leibniz Universität Hannover, Callinstrasse 38, D-30167, Hannover, Germany*

The stochastic gravitational wave background for pulsar timing arrays is often modeled by a Gaussian ensemble which is isotropic and unpolarized. However, the Universe has a discrete set of polarized gravitational wave sources at specific sky locations. Can we trust that the Gaussian ensemble is an accurate description? To investigate this, we explicitly construct an ensemble containing $N$ individual binary sources with circular orbits. The orbital inclination angles are randomly distributed, hence the individual sources are elliptically polarized. We then compute the first two moments of the Hellings and Downs correlation, as well as the pulsar-averaged correlation mean and (cosmic) variance. The first moments are the same as for a previously studied ensemble of circularly polarized sources. However, the second moments, and hence the variances, are different for the two ensembles. While neither discrete source model is exactly described by a Gaussian ensemble, we show that in the limit of large $N$, the differences are small.

## I. INTRODUCTION

Pulsar timing arrays (PTAs) appear to be on the verge of making the first detections [1–4] of nano-Hertz gravitational waves (GWs). Much of the data analysis and tools [5] assume that the GWs are described by an isotropic and unpolarized Gaussian ensemble [6–8]. However, this is an approximation: a realistic collection of discrete sources is only described by the Gaussian ensemble in the limit of an infinite density of weak sources [9]. This paper investigates the differences.

The most likely PTA GW sources [10] are gravitationally bound binary systems consisting of two black holes with masses $10^4$–$10^{10} M_\odot$, where $M_\odot$ is the solar mass. In the simplest source models, which we adopt here, the orbits are assumed to be circular.

The polarization of the GWs produced by a circular binary depend upon the angle of inclination $\iota$. This is the angle between the line of sight and the orbital angular momentum. The GWs which are produced are linearly polarized if $\iota = 90°$ (orbit seen edge-on). The GWs are circularly polarized if $\iota = 0°$ or $\iota = 180°$ (orbit seen face-on or face-off). For other values of $\iota$, the GWs are elliptically polarized. This begs the question, would an ensemble of such elliptically polarized sources produce a GW background with the same properties as an ensemble of unpolarized sources? (Note that the language can be misleading. A source is called "unpolarized" if (a) the time average of $(h^+)^2 - (h^\times)^2$ vanishes and (b) the time average of $h^+ h^\times$ vanishes; see the paragraph following Eq. (24) of [11]. Thus, a circularly polarized source is "unpolarized.")

Reference [9] constructs a GW background by summing the GWs produced by a discrete set of unpolarized sources: the sources are circularly polarized, corresponding to $\iota = 0°$ or $\iota = 180°$. Here, we improve this simple GW model, assuming instead that $\cos \iota$ is uniformly distributed on the interval $[-1, 1]$. This corresponds to a physically realistic random distribution of orbital inclination angles. To characterize the resulting GW background, we compute the mean and the variance of the Hellings and Downs (HD) correlation.

A brief outline of the paper is as follows. We begin in Sec. II with the GW amplitude produced by a very distant circular binary system labeled by an integer $j$. In Sec. III, we show how to treat the parameters as random variables, and explain the relationship to the simpler ensemble of face-on and face-off binary systems presented in [9]. In Sec. IV, as a "warm-up exercise," we compute the first and second moments and variance of a local quantity: the time-averaged squared strain $s$ defined on the first line of (4.6). We then compare this quantity for the ensemble of unpolarized sources from [9] and our ensemble of polarized sources. In Sec. V, we repeat these calculations of the mean, second moment, and variance for the HD correlation between two pulsars. Again, we compare it for the two ensembles. In Sec. VI, we define a pulsar-averaged correlation at angle $\gamma$, and compute its mean, second moment, and (cosmic) variance, and compare it for the two ensembles. In Sec. VII, the results are compared to those of the standard Gaussian ensemble, and we show that in a certain $N \to \infty$ limit our results approach those of the Gaussian ensemble. This is followed by a short conclusion.

[*]bruce.allen@aei.mpg.de
[†]serena.valtolina@aei.mpg.de

We use units in which the speed of light is unity, meaning that distances are expressed in terms of the time that it would take light or GWs to propagate that distance.

## II. GW SOURCE MODEL

Here, we follow the approach of Sec. 3A of [9]. In a representative universe, the GW background is produced by a discrete collection of $N$ distant point sources. The sources are indexed by $j = 1, ..., N$, where $j = 1$ is the closest source, $j = 2$ is the second closest, and so on. The GWs arrive at Earth from the $j$th source with propagation direction $\mathbf{\Omega}_j$. Hence, the line of sight to each source is the unit-length vector $-\mathbf{\Omega}_j$, though we often call $\mathbf{\Omega}_j$ the "source direction."

In the model of [9], each source has circularly polarized gravitational waveforms

$$h_j^+(t) = \mathcal{A}_j \cos(\omega t + \phi_j),$$
$$h_j^\times(t) = \mathcal{A}_j \sin(\omega t + \phi_j), \quad (2.1)$$

where $h^+$ and $h^\times$ are the GW amplitudes in the respective polarizations at Earth at time $t$. Here, $t$ is time at the solar system barycenter (SSB) [12], and $\omega$ is an angular frequency. The amplitudes $\mathcal{A}_j \equiv \mathcal{A}/j^{1/3}$ fall off with distance, corresponding to a uniform density in 3-dimensional Euclidean space (out to some maximum radius). The constant $\mathcal{A}$ is the GW amplitude of the closest source, measured at the SSB.

The orbital frequency of a binary system increases with time as the system radiates GWs and loses energy, becoming more tightly bound. Here, we assume that this orbital frequency is evolving slowly on the time scale of PTA observations. Thus, the GW frequency $\omega$ (nominally, twice the orbital frequency) may be treated as a constant. We also assume that $\omega$ is commensurate with the total observation time $T$, meaning that $\omega T/2\pi$ is integer. Note that, in this model, all $N$ sources emit GWs at the same frequency $\omega$. Since the frequency is commensurate with $T$, they all fall into a single frequency bin.

In the vicinity of the PTA, far from the emitting binary systems, the GWs are weak, so the total GW field may be written as the sum of the GWs from the individual sources:

$$h_{ab}(t, \mathbf{x}) \equiv \sum_{j=1}^{N} h_{j,ab}(t, \mathbf{x}). \quad (2.2)$$

The GW emitted by the $j$th source is

$$h_{j,ab}(t, \mathbf{x}) \equiv h_j^+(t - \mathbf{x} \cdot \mathbf{\Omega}_j) e_{ab}^+(\mathbf{\Omega}_j) + h_j^\times(t - \mathbf{x} \cdot \mathbf{\Omega}_j) e_{ab}^\times(\mathbf{\Omega}_j), \quad (2.3)$$

where $+$ and $\times$ label the two GW polarization states, and the traceless and symmetric polarization tensors $e_{ab}^+$ and $e_{ab}^\times$ are defined in Eq. (D6) [9]. These polarization states are defined with respect to the propagation direction $\mathbf{\Omega}_j$ of the GWs arriving from the $j$th point source. The definition uses an arbitrary direction orthogonal to $\mathbf{\Omega}$; our choice is that of Eqs. (D6) and (D7) of [9]. In the model of (2.1), the $+$ and $\times$ waveforms have equal amplitudes, but are shifted in phase by 90°. This is because [9] assumes, for simplicity, that each of the individual point sources is circularly polarized. In what follows, for convenience, we drop the limits from the summation if they are $j = 1$ to $N$ as in (2.2).

Note that these expressions are only valid in the region of space containing Earth and the pulsars, a few thousand years in size. In that neighborhood they are a good approximation: the sources are so distant that the deviations between their spherical wavefronts and plane waves are much smaller than a gravitational wavelength.

A binary system in a circular orbit with an inclination angle $\iota$ in the range $(0°, 180°)$ does not produce circularly polarized GWs. It produces an elliptical polarization, with amplitudes [13–15]

$$h_j^+(t) = A_j \left[\frac{1}{2}(1 + \cos^2 \iota_j) \cos 2\psi_j \cos(\omega t + \phi_j) \right.$$
$$\left. - \cos \iota_j \sin 2\psi_j \sin(\omega t + \phi_j)\right],$$
$$h_j^\times(t) = A_j \left[\frac{1}{2}(1 + \cos^2 \iota_j) \sin 2\psi_j \cos(\omega t + \phi_j) \right.$$
$$\left. + \cos \iota_j \cos 2\psi_j \sin(\omega t + \phi_j)\right], \quad (2.4)$$

which should be substituted into (2.3). As in (2.1), the GW amplitudes $A_j \equiv A/j^{1/3}$ correspond to a constant source density within a ball, where $A$ is the amplitude of the closest source. Later, we will see that the two different ensembles have the same expected squared GW strain at Earth if $\mathcal{A} = \sqrt{2/5}A$.

Equation (2.4) is similar to (2.1), but contains two additional parameters $\iota_j$ and $\psi_j$, which describe aspects of the binary orbit. As previously explained, $\iota \in [0°, 180°]$ is the angle between the line of sight and a vector normal to the orbital plane. For $\iota = 0°$ or $\iota = 180°$, the orbit appears circular when viewed from Earth (i.e., projected on the plane of the sky). For any other orbital inclination, the orbit appears to be a noncircular ellipse. The parameter $\psi_j \in [0°, 90°]$ describes the orientation of this ellipse. It is the angle between two vectors, after projection into a plane orthogonal to $\mathbf{\Omega}_j$. The first points along the major (long) axis of the orbital ellipse, as seen from Earth. The second is **m**, as defined in Eq. (D7) of [9].

Often the parameter $\psi$ does not appear. Indeed, if there is only one binary system, then $\psi$ can be set to zero in (2.4). This follows from gauge invariance: $\psi$ can be eliminated by selecting polarization axes which are aligned with the ellipse. More precisely, the $+$ and $\times$ polarization directions are defined by a pair of vectors **m** and **n**, which are orthogonal to the GW propagation direction $\boldsymbol{\Omega}$. If there is only a single source, then picking the direction **m** along the line of the major axis of the ellipse is equivalent to setting $\psi_j = 0$. If there is more than one source, then one could define vectors **m**' and **n**' on a per-source basis, to eliminate $\psi_j$. However, if we do this, then the calculations that follow would contain new source-dependent quantities $\mathbf{m}'_j$ and $\mathbf{n}'_j$; the number of variables and corresponding complexity is not reduced.

## III. GW ENSEMBLE

To compute statistical properties of the GW background, we consider an ensemble of different universes, each containing $N$ sources. In each of these universes (also called "realizations"), (2.4) defines the GW emission from the $j$th source. Thus, each realization is defined by particular values of $\phi_j$, $\boldsymbol{\Omega}_j$, $\iota_j$, and $\psi_j$, for $j = 1, \ldots, N$. But, since we are ignorant of the values that these quantities have in our own universe, we must resort to statistical inferences about "typical" members of the ensemble. For this purpose, we treat these quantities as random variables, and use ensemble averages to predict the properties of a typical member of the ensemble.

For binary systems with random orientations and positions in space, the $5N$ random variables are independent. The orbital phases $\phi_j$ are uniformly distributed over $[0, 2\pi]$. The directions to the sources $\boldsymbol{\Omega}_j$ are uniformly distributed over the two-sphere. The cosines of $\iota_j$ are uniformly distributed on $[-1, 1]$, and the polarization angles $\psi_j$ are uniformly distributed over $[0, \pi/2]$.

The correct way to describe this ensemble is as "an ensemble of polarized sources," since each realization of the ensemble contains many elliptically polarized GW sources (2.4). However, we frequently employ the shorthand "the polarized ensemble" or (for the contrasting case) "the unpolarized ensemble." [The circularly polarized GW sources in the latter are described by (2.1).] The shorthand "polarized ensemble" can be misleading, because the ensemble itself has no preferred direction or axis or polarization. The reader should bear in mind that it is not the *ensemble* which is polarized or unpolarized, but rather the individual sources in each realization of that ensemble.

## IV. SQUARED-STRAIN ESTIMATOR s: MEAN, SECOND MOMENT, AND VARIANCE

Following [9,16] we first compute the expected value and variance of the time-averaged squared strain. The calculation is similar to that used to find the mean and variance of the HD correlation, but is simpler.

It is helpful to introduce a complex polarization basis

$$e_{j,ab} = e_{ab}(\boldsymbol{\Omega}_j) \equiv e^+_{ab}(\boldsymbol{\Omega}_j) - i e^\times_{ab}(\boldsymbol{\Omega}_j), \quad (4.1)$$

where $j$ labels the source and $a$ and $b$ are tangent-space indices. $e_{ab}(\boldsymbol{\Omega}_j)$ contains both the plus and cross components, and corresponds to a left-circularly-polarized GW. The complex conjugate $e^*_{j,ab}$ is the polarization of a right-circularly-polarized GW.

These complex polarization tensors form a null basis for a two-dimensional vector space. From the definitions of $e^+_{ab}$ and $e^\times_{ab}$ given in Eqs. (D6) and (D7) of [9], it follows that $e^+_{ab}e^{+ab} = e^\times_{ab}e^{\times ab} = 2$. This implies that

$$\begin{aligned} e_{ab}e^{*ab} &= 4, \\ e_{ab}e^{ab} &= 0, \\ e^*_{ab}e^{*ab} &= 0, \end{aligned} \quad (4.2)$$

where the third line follows from the second one.

It is also convenient to introduce several functions that will simplify the notation. We define

$$\begin{aligned} U_j &\equiv A_j(1 + \cos \iota_j)^2, \\ V_j &\equiv A_j(1 - \cos \iota_j)^2, \end{aligned} \quad (4.3)$$

which only depend upon the orbital inclination angle. Using these, we also define

$$\begin{aligned} \mathcal{U}_j &\equiv U_j \, e^{2i\psi_j}, \\ \mathcal{V}_j &\equiv V_j \, e^{2i\psi_j}, \end{aligned} \quad (4.4)$$

which depend on both the orbital inclination and on the polarization angle.

We now compute the time average of the squared GW amplitude at Earth ($\mathbf{x} = 0$). First, we substitute (4.3) and (4.4) into (2.4). The resulting expression is inserted into (2.3), and (4.1) and its complex conjugate are used to replace $e^+_{ab}$ and $e^\times_{ab}$. The GW strain at Earth becomes

$$h_{ab}(t,0) = \frac{1}{8} \sum_j \left[ (\mathcal{U}_j e_{j,ab} + \mathcal{V}^*_j e^*_{j,ab}) e^{i(\omega t + \phi_j)} \right.$$

$$\left. + (\mathcal{U}^*_j e^*_{j,ab} + \mathcal{V}_j e_{j,ab}) e^{-i(\omega t + \phi_j)} \right]. \quad (4.5)$$

We now square and contract this quantity, and time-average, obtaining

$$s \equiv \overline{h_{ab}(t,0)h^{ab}(t,0)}$$
$$= \frac{1}{32}\sum_j \left[(\mathcal{U}_j e_{j,ab} + \mathcal{V}_j^* e_{j,ab}^*)(\mathcal{U}_j^* e_j^{*ab} + \mathcal{V}_j e_j^{ab})\right]$$
$$+ \frac{1}{64}\sum_{j\neq k}\left[(\mathcal{U}_j e_{j,ab} + \mathcal{V}_j^* e_{j,ab}^*)(\mathcal{U}_k^* e_k^{*ab} + \mathcal{V}_k e_k^{ab})e^{i(\phi_j - \phi_k)}\right.$$
$$\left. + (\mathcal{U}_j^* e_{j,ab}^* + \mathcal{V}_j e_{j,ab})(\mathcal{U}_k e_k^{ab} + \mathcal{V}_k^* e_k^{*ab})e^{-i(\phi_j - \phi_k)}\right]. \quad (4.6)$$

The average over time $\bar{Q} \equiv \frac{1}{T}\int_{-T/2}^{T/2} Q(t)dt$ has eliminated half of the cross terms, and we have grouped the others into "diagonal" terms with $j = k$ and "off-diagonal" terms with $j \neq k$.

In this paper, we use different indexing and summation conventions for tangent-space indices such as $a$, $b$, $c$, and $d$, and for source labels such as $j$, $k$, $\ell$, and $m$. We adopt the Einstein summation convention for repeated tangent-space indices. For example, in (4.6) the indices $a$ and $b$ are implicitly summed from 1 to 3. In contrast, sums over GW sources are explicit. Double sums over sources are indicated with a single summation symbol as $\sum_{j,k}$. If the terms with $j = k$ are dropped, then the double sums are denoted by $\sum_{j\neq k}$.

Expression (4.6) for $s$ holds for any realization in the ensemble, and is a function of the $5N$ variables $\phi_j$, $\mathbf{\Omega}_j$, $\iota_j$, and $\psi_j$. For a given ensemble, we can calculate the expected value of *any* function of those $5N$ variables, including $s$ or powers of $s$ or other functions of $s$.

If $Q$ is a function or functional of the $5N$ variables, then the ensemble average of $Q$ is defined by

$$\langle Q \rangle \equiv \prod_{j=1}^N \left(\int_0^{2\pi} \frac{d\phi_j}{2\pi}\right) \prod_{k=1}^N \left(\int_0^{2\pi} \frac{d\psi_k}{2\pi}\right) \prod_{\ell=1}^N \left(\int_{-1}^1 \frac{d\cos\iota_\ell}{2}\right)$$
$$\times \prod_{m=1}^N \left(\int \frac{d\mathbf{\Omega}_m}{4\pi}\right) Q(\phi_1, \ldots, \mathbf{\Omega}_N). \quad (4.7)$$

In practice, we carry out these integrals in four stages. If we first carry out the $N$ integrals over $\phi_j$ in (4.7), the result may be denoted by $\langle Q \rangle_\phi$. Doing the remaining three sets of integrals then gives the full average. Thus, $\langle Q \rangle \equiv \langle\langle\langle\langle Q \rangle_\phi \rangle_\psi \rangle_\iota \rangle_\mathbf{\Omega}$, where each of the separate averages refers to one set of the integrals in (4.7) [17].

### A. First moment of $s$

The first moment of $s$ is easily computed. It follows immediately from (4.7) that $\langle e^{i(\phi_j - \phi_k)}\rangle_\phi$ vanishes if $j \neq k$ and is unity if $j = k$. Hence, if we take the $\phi_j$ average of (4.6), only the diagonal term survives, giving

$$\langle s \rangle_\phi = \frac{1}{32}\sum_j \left[(\mathcal{U}_j e_{j,ab} + \mathcal{V}_j^* e_{j,ab}^*)(\mathcal{U}_j^* e_j^{*ab} + \mathcal{V}_j e_j^{ab})\right]. \quad (4.8)$$

We next average over $\psi_j$, carrying out the next set of integrals in (4.7). The rhs of (4.8) only depends upon $\psi_j$ through $\mathcal{U}_j$ and $\mathcal{V}_j$, as defined by (4.4). The $\mathcal{U}_j\mathcal{V}_j$ and $\mathcal{U}_j^*\mathcal{V}_j^*$ terms are proportional to $\exp(\pm 4i\psi_j)$ and integrate to zero. The surviving $\mathcal{U}_j\mathcal{U}_j^*$ and $\mathcal{V}_j^*\mathcal{V}_j$ terms give

$$\langle\langle s \rangle_\phi\rangle_\psi = \frac{1}{32}\sum_j (U_j^2 + V_j^2) e_{j,ab}^* e_j^{ab}. \quad (4.9)$$

From (4.3), one can see that the factor $U_j^2 + V_j^2$ only depends upon the inclination angle $\iota_j$. Evaluating the third set of integrals in (4.7) to average over $\iota_j$ gives

$$\langle U_j^2 + V_j^2 \rangle_\iota = \frac{1}{2}A_j^2 \int_{-1}^1 dx\,((1+x)^4 + (1-x)^4)$$
$$= \frac{32}{5}A_j^2, \quad (4.10)$$

where $x = \cos\iota_j$. Thus, the $\iota$-average of (4.9) is

$$\langle\langle\langle s \rangle_\phi\rangle_\psi\rangle_\iota = \frac{1}{5}\sum_j A_j^2 e_{j,ab}^* e_j^{ab}. \quad (4.11)$$

The final quantities to average are the polarization tensors, which only depend upon the direction $\mathbf{\Omega}_j$. In fact, it follows immediately from their definition (see Eqs. (D6) and (D7) of [9]) that $e_{ab}^*(\mathbf{\Omega})e^{ab}(\mathbf{\Omega}) = 4$ is independent of $\mathbf{\Omega}$ (so the label "$j$" is not needed). Hence, from (4.11) we obtain the ensemble average

$$\langle s \rangle_{\text{pol}} = \frac{4}{5}\sum_j A_j^2. \quad (4.12)$$

This completes the calculation of the first moment for the polarized ensemble (2.4).

### B. Second moment and variance of $s$

We next compute the second moment of the time-averaged squared strain $s$. We will use this to determine the variance of $s$. To evaluate $\langle s^2 \rangle$, we square $s$ as given in (4.6) and then use (4.7) to determine the ensemble average. This is very similar to the calculation of the first moment, so we can be more concise.

We begin by computing the average over the phases $\phi_j$. First, square (4.6) to obtain three terms. These are (i) the square of the diagonal sum, (ii) the square of the off-diagonal sum, and (iii) twice their cross-product. Averaging over the random phases $\phi_j$ eliminates (iii), and the remaining terms are

$$\langle s^2 \rangle_\phi = \left( \frac{1}{32} \sum_j \left[ (\mathcal{U}_j e_{j,ab} + \mathcal{V}_j^* e_{j,ab}^*)(\mathcal{U}_j^* e_j^{*ab} + \mathcal{V}_j e_j^{ab}) \right] \right)^2$$

$$+ \frac{1}{4096} \sum_{j \neq k} \sum_{\ell \neq m} \Big[ (\mathcal{U}_j e_{j,ab} + \mathcal{V}_j^* e_{j,ab}^*)(\mathcal{U}_k^* e_k^{*ab} + \mathcal{V}_k e_k^{ab})(\mathcal{U}_\ell e_{\ell,cd} + \mathcal{V}_\ell^* e_{\ell,cd}^*)(\mathcal{U}_m^* e_m^{*cd} + \mathcal{V}_m e_m^{cd}) \langle e^{i(\phi_j - \phi_k + \phi_\ell - \phi_m)} \rangle_\phi$$

$$+ (\mathcal{U}_j e_{j,ab} + \mathcal{V}_j^* e_{j,ab}^*)(\mathcal{U}_k^* e_k^{*ab} + \mathcal{V}_k e_k^{ab})(\mathcal{U}_\ell^* e_{\ell,cd}^* + \mathcal{V}_\ell e_{\ell,cd})(\mathcal{U}_m e_m^{cd} + \mathcal{V}_m^* e_m^{*cd}) \langle e^{i(\phi_j - \phi_k - \phi_\ell + \phi_m)} \rangle_\phi$$

$$+ (\mathcal{U}_j^* e_{j,ab}^* + \mathcal{V}_j e_{j,ab})(\mathcal{U}_k e_k^{ab} + \mathcal{V}_k^* e_k^{*ab})(\mathcal{U}_\ell e_{\ell,cd} + \mathcal{V}_\ell^* e_{\ell,cd}^*)(\mathcal{U}_m^* e_m^{*cd} + \mathcal{V}_m e_m^{cd}) \langle e^{-i(\phi_j - \phi_k - \phi_\ell + \phi_m)} \rangle_\phi$$

$$+ (\mathcal{U}_j^* e_{j,ab}^* + \mathcal{V}_j e_{j,ab})(\mathcal{U}_k e_k^{ab} + \mathcal{V}_k^* e_k^{*ab})(\mathcal{U}_\ell^* e_{\ell,cd}^* + \mathcal{V}_\ell e_{\ell,cd})(\mathcal{U}_m e_m^{cd} + \mathcal{V}_m^* e_m^{*cd}) \langle e^{-i(\phi_j - \phi_k + \phi_\ell - \phi_m)} \rangle_\phi \Big]. \quad (4.13)$$

The averages of the exponentials are unity if the exponent vanishes, else they are zero. This implies

$$\langle e^{\pm i(\phi_j - \phi_k + \phi_\ell - \phi_m)} \rangle_\phi = \delta_{jm} \delta_{k\ell},$$
$$\langle e^{\pm i(\phi_j - \phi_k - \phi_\ell + \phi_m)} \rangle_\phi = \delta_{j\ell} \delta_{km}, \quad (4.14)$$

provided that $j \neq k$ and $\ell \neq m$. Substituting (4.14) into (4.13) eliminates one of the double sums. The remaining double sum has four terms, which are all equal. Finally, expressing the square of the single sum as diagonal and off-diagonal terms gives

$$\langle s^2 \rangle_\phi = \frac{1}{1024} \sum_j \left[ (\mathcal{U}_j e_{j,ab} + \mathcal{V}_j^* e_{j,ab}^*)(\mathcal{U}_j^* e_j^{*ab} + \mathcal{V}_j e_j^{ab}) \right]^2$$

$$+ \frac{1}{1024} \sum_{j \neq k} \Big[ (\mathcal{U}_j e_{j,ab} + \mathcal{V}_j^* e_{j,ab}^*)(\mathcal{U}_j^* e_j^{*ab} + \mathcal{V}_j e_j^{ab})(\mathcal{U}_k e_{k,cd} + \mathcal{V}_k^* e_{k,cd}^*)(\mathcal{U}_k^* e_k^{*cd} + \mathcal{V}_k e_k^{cd})$$

$$+ (\mathcal{U}_j e_{j,ab} + \mathcal{V}_j^* e_{j,ab}^*)(\mathcal{U}_k^* e_k^{*ab} + \mathcal{V}_k e_k^{ab})(\mathcal{U}_k e_k^{cd} + \mathcal{V}_k^* e_k^{*cd})(\mathcal{U}_j^* e_{j,cd}^* + \mathcal{V}_j e_{j,cd}) \Big]. \quad (4.15)$$

This completes the averaging over $\phi_j$.

Next, we average (4.15) over $\psi_j$. From (4.4) we see that both $\mathcal{U}$ and $\mathcal{V}$ depend on $\psi$ only through the factor $\exp(2i\psi)$. Thus, after averaging over $\psi$, terms with different numbers of conjugated and unconjugated $\mathcal{U}$ and $\mathcal{V}$ will vanish. For example, $\mathcal{U}\mathcal{U}^*\mathcal{V}^2$ has one conjugated and three unconjugated; it averages to zero. The terms inside the single sum of (4.15) that will survive the averaging are

$$(\mathcal{U}_j^2 \mathcal{U}_j^{*2} + \mathcal{V}_j^2 \mathcal{V}_j^{*2})(e_j^{ab} e_{j,ab}^*)^2 + 2(\mathcal{U}_j \mathcal{U}_j^* \mathcal{V}_j \mathcal{V}_j^*) \Big( (e_j^{ab} e_{j,ab}^*)^2 + e_j^{ab} e_{j,ab} e_j^{*cd} e_{j,cd}^* \Big).$$

The terms inside the double sum of (4.15) that will survive the averaging are

$$(\mathcal{U}_j \mathcal{U}_j^* + \mathcal{V}_j \mathcal{V}_j^*) e_j^{ab} e_{j,ab}^* (\mathcal{U}_k \mathcal{U}_k^* + \mathcal{V}_k \mathcal{V}_k^*) e_k^{cd} e_{k,cd}^* + (\mathcal{U}_j \mathcal{U}_j^* e_{j,ab} e_{j,cd}^* + \mathcal{V}_j \mathcal{V}_j^* e_{j,cd} e_{j,ab}^*)(\mathcal{U}_k \mathcal{U}_k^* e_k^{ab} e_k^{*cd} + \mathcal{V}_k \mathcal{V}_k^* e_k^{cd} e_k^{*ab}).$$

Combining these, using (4.2) to evaluate the contractions of the polarization tensors, and making use of (4.4), the $\psi$-average of (4.15) becomes

$$\langle \langle s^2 \rangle_\phi \rangle_\psi = \frac{1}{64} \sum_j \left[ U_j^4 + V_j^4 + 2U_j^2 V_j^2 \right]$$

$$+ \frac{1}{1024} \sum_{j \neq k} \left[ 16(U_j^2 + V_j^2)(U_k^2 + V_k^2) + (U_j^2 e_{j,ab} e_{j,cd}^* + V_j^2 e_{j,ab}^* e_{j,cd})(U_k^2 e_k^{*ab} e_k^{cd} + V_k^2 e_k^{ab} e_k^{*cd}) \right]. \quad (4.16)$$

Note that the condition $j \neq k$ simplifies the evaluation of the $\psi$-average of the double sum: the average of the product becomes the product of the averages.

We next average over $\iota_j$, as defined in (4.7). From (4.3), the only $\iota$ dependence is through the functions $U$ and $V$. For example,

$$\langle U_j^2 \rangle_\iota = \frac{1}{2}\int_{-1}^{1} U_j^2 \, d\cos\iota_j = \frac{1}{2}A_j^2 \int_{-1}^{1}(1+x)^4 dx = \frac{16}{5}A_j^2, \quad (4.17)$$

where $x = \cos\iota_j$. For this calculation, the following averages, computed in the same way, are sufficient:

$$\langle U_j^2 \rangle_\iota = \langle V_j^2 \rangle_\iota = \frac{16}{5}A_j^2,$$
$$\langle U_j^4 \rangle_\iota = \langle V_j^4 \rangle_\iota = \frac{256}{9}A_j^4,$$
$$\langle U_j^2 V_j^2 \rangle_\iota = \frac{128}{315}A_j^4. \quad (4.18)$$

Using (4.18) to average (4.16) over $\iota$ gives

$$\langle\langle\langle s^2\rangle_\phi\rangle_\psi\rangle_\iota = \frac{284}{315}\sum_j A_j^4 + \frac{1}{100}\sum_{j\neq k} A_j^2 A_k^2 \Big[64 + (e_{j,ab}e^*_{j,cd}$$
$$+ e^*_{j,ab}e_{j,cd})(e_k^{ab}e_k^{*cd} + e_k^{*ab}e_k^{cd})\Big]. \quad (4.19)$$

Again, because $j \neq k$, we write the average of the product of $j$- and $k$-dependent terms as the product of the averages.

The final step to compute the second moment is to average (4.19) over the source directions $\Omega_j$. For this purpose, we employ the spherical average of polarization tensors

$$\eta_{abcd} \equiv \frac{1}{4\pi}\int e_{ab}(\Omega)e^*_{cd}(\Omega)d\Omega$$
$$= -\frac{4}{15}\delta_{ab}\delta_{cd} + \frac{2}{5}(\delta_{ac}\delta_{bd} + \delta_{ad}\delta_{bc}), \quad (4.20)$$

which is derived in Eqs. (C33) and (C34) of [9]. Here, $\delta_{ab}$ denotes the 3-dimensional Kronecker delta. Using (4.20) to average (4.19) over the source directions $\Omega_j$ gives

$$\langle s^2 \rangle = \frac{284}{315}\sum_j A_j^4 + \frac{1}{25}\sum_{j\neq k} A_j^2 A_k^2(16 + \eta_{abcd}\eta^{abcd})$$
$$= \frac{284}{315}\sum_j A_j^4 + \frac{96}{125}\sum_{j\neq k} A_j^2 A_k^2$$
$$= \frac{1052}{7875}\sum_j A_j^4 + \frac{96}{125}\left(\sum_j A_j^2\right)^2. \quad (4.21)$$

Since $j \neq k$, the first line follows from carrying out the final averaging in (4.7) *separately* on the $j$ and the $k$ terms. To obtain the second line, we use $\eta_{abcd}\eta^{abcd} = 16/5$, obtained from (4.20). The third line follows from

$$\sum_{j\neq k} A_j^2 A_k^2 = \left(\sum_j A_j^2\right)^2 - \sum_j A_j^4 \quad (4.22)$$

(which holds for *any* set of amplitudes). This completes the calculation of the second moment of $s$.

The variance $\sigma_s^2 \equiv \langle s^2 \rangle - \langle s \rangle^2$ is obtained from the first (4.12) and second (4.21) moments. The mean and variance of the polarized ensemble (2.4) are

$$\langle s \rangle_{\text{pol}} = \frac{4}{5}\sum_j A_j^2,$$
$$\sigma_{s,\text{pol}}^2 = \frac{1052}{7875}\sum_j A_j^4 + \frac{16}{125}\left(\sum_j A_j^2\right)^2, \quad (4.23)$$

where the sum is over the $N$ sources, with amplitudes given after (2.4). We now compare this mean and variance to those of the unpolarized ensemble (2.1).

## C. Comparison of polarized and unpolarized ensembles

The mean and variance of $s$ for the unpolarized ensemble (2.1) are given in Eqs. (3.15) and (3.23) of [9] as

$$\langle s \rangle_{\text{unpol}} = 2\sum_j \mathcal{A}_j^2,$$
$$\sigma_{s,\text{unpol}}^2 = -\frac{4}{5}\sum_j \mathcal{A}_j^4 + \frac{4}{5}\left(\sum_j \mathcal{A}_j^2\right)^2. \quad (4.24)$$

Both ensembles are characterized by a single number: the amplitude of the closest source to Earth, denoted respectively by $A$ and $\mathcal{A}$.

In order to compare the two ensembles "on equal footing," we select these amplitudes so that both ensembles have the same mean squared GW amplitude $\langle s \rangle$ at Earth. Comparing (4.23) and (4.24), this implies that

$$\mathcal{A}^2 = \frac{2}{5}A^2 \Leftrightarrow \mathcal{A}_j = \sqrt{\frac{2}{5}}A_j. \quad (4.25)$$

It follows immediately from (4.23) and (4.24) that for a given mean, the variance of the polarized ensemble is always larger than that of the unpolarized ensemble.

These expressions include $N$ sources labeled by $j = 1,\ldots N$. Because $A_j$ and $\mathcal{A}_j$ fall off $\propto j^{-1/3}$, the sums $\sum_j A_j^2$ and $\sum_j \mathcal{A}_j^2$ diverge as the number of GW sources $N \to \infty$. In contrast, the sums $\sum_j A_j^4$ and $\sum_j \mathcal{A}_j^4$ converge to $A^4\zeta(4/3) \approx 3.6A^4$ and $\mathcal{A}^4\zeta(4/3) \approx 3.6\mathcal{A}^4$ respectively, where $\zeta$ is the Riemann zeta function. Thus, in the limit of many sources, the fractional difference in the variances approaches zero.

## V. MEAN AND VARIANCE OF THE HD CORRELATION

The HD correlation $\rho$ between a pair of pulsars is defined in terms of their redshifts $Z$. Consider a pulsar at spatial location $\mathbf{x} = L\hat{\mathbf{p}}$, where $L$ is the light-travel time to the pulsar and $\hat{\mathbf{p}}$ is a unit-length vector. The redshift induced by the $j$th source at time $t$ is

$$Z_j(t) \equiv \frac{1}{2} \frac{\hat{p}^a \hat{p}^b}{1+\mathbf{\Omega}_j \cdot \hat{\mathbf{p}}} \left[ h_{j,ab}(t,0) - \chi h_{j,ab}(t-L, L\hat{\mathbf{p}}) \right], \quad (5.1)$$

where the terms in square brackets are called the "Earth term" and the "pulsar term" respectively. For a PTA, the correct value of $\chi$ is $\chi = 1$, but we retain this dependence to better understand the effects of the pulsar term. The GW amplitudes $h_{j,ab}(t,\mathbf{x})$ are defined in (2.3) as a linear combination of the two polarizations $e_{ab}^+(\mathbf{\Omega}_j)$ and $e_{ab}^\times(\mathbf{\Omega}_j)$.

The GW amplitude of the $j$th source $h_{j,ab}(t,\mathbf{x})$ may also be expressed in other bases, such as the complex polarization basis $e_{j,ab}$ introduced in (4.1). For what follows, it is convenient to introduce the rotated complex basis

$$e^{2i\psi_j} e_{j,ab}. \quad (5.2)$$

We can then write $h_{j,ab}(t,\mathbf{x})$ as a linear combination of (5.2) and its complex conjugate. Defining the complex GW amplitude

$$h_j(t) \equiv \frac{1}{2} A_j \Big[ (1+\cos^2\iota_j)\cos(\omega t + \phi_j) \\ + 2i\cos\iota_j \sin(\omega t + \phi_j) \Big], \quad (5.3)$$

it follows from (2.3), (2.4) and (4.1) that

$$h_{j,ab}(t,\mathbf{x}) = \frac{1}{2} \Big[ h_j(t-\mathbf{x}\cdot\mathbf{\Omega}_j) e^{2i\psi_j} e_{j,ab} \\ + h_j^*(t-\mathbf{x}\cdot\mathbf{\Omega}_j) e^{-2i\psi_j} e_{j,ab}^* \Big]. \quad (5.4)$$

This can be substituted into (5.1) to determine the redshift $Z$. In effect, for a given GW propagation direction, we have replaced the two degrees of freedom contained in $h_{j,ab}$ (the amplitudes of the cross and plus polarizations) with a single complex amplitude $h_j$.

Since the induced redshift (5.1) depends upon the difference between Earth $(t, \mathbf{x}=0)$ and pulsar $(t-L, \mathbf{x}=L\hat{\mathbf{p}})$ terms, we introduce the differences

$$\Delta h_j(t) \equiv h_j(\tau - \mathbf{x}\cdot\mathbf{\Omega}_j)|_{\tau=t,\mathbf{x}=0} - h_j(\tau - \mathbf{x}\cdot\mathbf{\Omega}_j)|_{\tau=t-L,\mathbf{x}=L\hat{\mathbf{p}}}$$
$$= \frac{1}{2} A_j(1+\cos^2\iota_j)\Big(\cos(\omega t + \phi_j) - \chi\cos(\omega t + \phi_j - \Delta_j)\Big) + iA_j\cos\iota_j\Big(\sin(\omega t + \phi_j) - \chi\sin(\omega t + \phi_j - \Delta_j)\Big)$$
$$= \frac{1}{4} U_j [1-\chi e^{-i\Delta_j}] e^{i(\omega t + \phi_j)} + \frac{1}{4} V_j [1-\chi e^{i\Delta_j}] e^{-i(\omega t + \phi_j)}. \quad (5.5)$$

The second line follows from (5.3) and the definition

$$\Delta_j \equiv \omega L (1 + \mathbf{\Omega}_j \cdot \hat{\mathbf{p}}), \quad (5.6)$$

and the third line from the definitions of $U_j$, $V_j$ given in (4.3). Starting with (5.1), and using (5.4) and the first line of (5.5), the induced redshift may be written as

$$Z_j(t) = \frac{1}{4} \frac{\hat{p}^a \hat{p}^b}{1+\mathbf{\Omega}_j \cdot \hat{\mathbf{p}}} \Big[ \Delta h_j(t) e^{2i\psi_j} e_{j,ab} + \Delta h_j^*(t) e^{-2i\psi_j} e_{j,ab}^* \Big]. \quad (5.7)$$

Using the final line of (5.5), this becomes

$$Z_j(t) = \frac{1}{16} \frac{\hat{p}^a \hat{p}^b}{1+\mathbf{\Omega}_j \cdot \hat{\mathbf{p}}} \Big[ \Big( U_j(1-\chi e^{-i\Delta_j}) e^{i(\omega t + \phi_j)} + V_j(1-\chi e^{i\Delta_j}) e^{-i(\omega t + \phi_j)} \Big) e^{2i\psi_j} e_{j,ab} \\ + \Big( U_j(1-\chi e^{i\Delta_j}) e^{-i(\omega t + \phi_j)} + V_j(1-\chi e^{-i\Delta_j}) e^{i(\omega t + \phi_j)} \Big) e^{-2i\psi_j} e_{j,ab}^* \Big]. \quad (5.8)$$

This expression for the redshift can be made more compact by writing it in terms of "antenna pattern" functions.

For a plane GW with direction $\mathbf{\Omega}$ and polarization $A = +, \times$, the (real) antenna pattern function for a given pulsar direction $\hat{\mathbf{p}}$ is defined as

$$F^A(\mathbf{\Omega}) \equiv \frac{1}{2} \frac{\hat{p}^a \hat{p}^b}{1+\mathbf{\Omega}\cdot\hat{\mathbf{p}}} e_{ab}^A(\mathbf{\Omega}). \quad (5.9)$$

The two polarizations can be combined into a single complex quantity by defining a complex antenna pattern function

$$F_j \equiv F^+(\mathbf{\Omega}_j) - iF^\times(\mathbf{\Omega}_j) = \frac{1}{2}\frac{\hat{p}^a \hat{p}^b}{1 + \mathbf{\Omega}_j \cdot \hat{\mathbf{p}}}\, e_{ab}(\mathbf{\Omega}_j), \quad (5.10)$$

where the second equality follows from the definition of the complex polarization tensor $e_{ab}$ in (4.1). If we use (5.10) to rewrite (5.8), then the only terms that depend upon the pulsar position $\hat{\mathbf{p}}$ are $F_j$ and $(1 - \chi e^{i\Delta_j})$, and their complex conjugates. Thus, we define

$$f_j \equiv 1 - \chi e^{i\Delta_j}, \quad (5.11)$$

where $\Delta_j$ is defined in (5.6) and depends upon the source direction $\mathbf{\Omega}_j$. (The term equal to one is the Earth term and the term proportional to $\chi$ is the pulsar term.) Using (5.10) and (5.11), we can write (5.8) as a linear combination of antenna pattern functions

$$Z_j(t) = \frac{1}{8}\left(U_j f_j^* e^{i(\omega t + \phi_j)} + V_j f_j e^{-i(\omega t + \phi_j)}\right) e^{2i\psi_j} F_j$$
$$+ \frac{1}{8}\left(U_j f_j e^{-i(\omega t + \phi_j)} + V_j f_j^* e^{i(\omega t + \phi_j)}\right) e^{-2i\psi_j} F_j^*,$$
$$(5.12)$$

which is the redshift induced by the $j$th source.

To compute the HD correlation, we need the total redshift induced by all $N$ GW sources. For weak gravitational fields, the redshifts add linearly. Thus, the total redshift at the time $t$ is the sum of the contributions (5.12) from each source:

$$Z_{\hat{\mathbf{p}}}(t) \equiv \sum_j Z_j(t)$$
$$= \frac{1}{8}\sum_j \Big[(U_j e^{2i\psi_j} F_j + V_j e^{-2i\psi_j} F_j^*) f_j^* e^{i(\omega t + \phi_j)}$$
$$+ (U_j e^{-2i\psi_j} F_j^* + V_j e^{2i\psi_j} F_j) f_j e^{-i(\omega t + \phi_j)}\Big].$$
$$(5.13)$$

We have regrouped the terms to isolate the positive- and negative-frequency terms. Note that the redshift depends upon the direction to the pulsar via the functions $f$ and $F$; we indicate this with the subscript $\hat{\mathbf{p}}$.

The correlation between two distinct pulsars with directions $\hat{\mathbf{p}}_1$ and $\hat{\mathbf{p}}_2$ is defined by

$$\rho \equiv \overline{Z_{\hat{\mathbf{p}}_1}(t) Z_{\hat{\mathbf{p}}_2}(t)}, \quad (5.14)$$

where the overbar [defined immediately after (4.6)] denotes a time average. We use (5.13) in (5.14) and average over time. Because $\omega$ is assumed to be an integer multiple of $2\pi/T$, only the cross terms between positive and negative frequency survive. We obtain

$$\rho = \frac{1}{64}\sum_{j,k}\Big[(U_j e^{2i\psi_j} F_{1,j} + V_j e^{-2i\psi_j} F_{1,j}^*)(U_k e^{-2i\psi_k} F_{2,k}^* + V_k e^{2i\psi_k} F_{2,k}) f_{1,j}^* f_{2,k} e^{i(\phi_j - \phi_k)}$$
$$+ (U_j e^{-2i\psi_j} F_{1,j}^* + V_j e^{2i\psi_j} F_{1,j})(U_k e^{2i\psi_k} F_{2,k} + V_k e^{-2i\psi_k} F_{2,k}^*) f_{1,j} f_{2,k}^* e^{-i(\phi_j - \phi_k)}\Big]. \quad (5.15)$$

The dependence of the antenna patterns $f$ and $F$ on pulsar directions $\hat{\mathbf{p}}_1$ and $\hat{\mathbf{p}}_2$ is denoted via the subscripts 1 and 2. We next compute the mean and variance of $\rho$ along the same lines as used for $s$ in Secs. IV A and IV B.

### A. First moment of $\rho$

We proceed as in Sec. IV A, evaluating the first set of integrals in (4.7) to average $\rho$ over the phases $\phi_j$. We obtain

$$\langle \rho \rangle_\phi = \frac{1}{64}\sum_j\Big[(U_j e^{2i\psi_j} F_{1,j} + V_j e^{-2i\psi_j} F_{1,j}^*)(U_j e^{-2i\psi_j} F_{2,j}^* + V_j e^{2i\psi_j} F_{2,j}) f_{1,j}^* f_{2,j}$$
$$+ (U_j e^{-2i\psi_j} F_{1,j}^* + V_j e^{2i\psi_j} F_{1,j})(U_j e^{2i\psi_j} F_{2,j} + V_j e^{-2i\psi_j} F_{2,j}^*) f_{1,j} f_{2,j}^*\Big], \quad (5.16)$$

where only the terms with $j = k$ remain.

We next average over $\psi$. Expanding the rhs of (5.16) gives terms proportional to $e^{4i\psi_j}$, $e^{-4i\psi_j}$, and $e^{0i\psi_j}$. Carrying out the second set of integrals in (4.7) eliminates the $e^{\pm 4i\psi}$ terms, leaving

$$\langle\langle \rho \rangle_\phi\rangle_\psi = \frac{1}{64}\sum_j\Big[(U_j^2 F_{1,j} F_{2,j}^* + V_j^2 F_{1,j}^* F_{2,j}) f_{1,j}^* f_{2,j} + (U_j^2 F_{1,j}^* F_{2,j} + V_j^2 F_{1,j} F_{2,j}^*) f_{1,j} f_{2,j}^*\Big]. \quad (5.17)$$

The next average is easy because all of the $\iota$ dependence is in $U$ and $V$.

The average over $\iota$ is defined by the third set of integrals in (4.7). Using $\langle U_j^2 \rangle_\iota = \langle V_j^2 \rangle_\iota = 16 A_j^2/5$ from (4.18) gives

$$\langle\langle\langle\rho\rangle_\phi\rangle_\psi\rangle_\iota = \frac{1}{20}\sum_j A_j^2 (F_{1,j}F_{2,j}^* + F_{1,j}^* F_{2,j})$$
$$\times (f_{1,j}^* f_{2,j} + f_{1,j}f_{2,j}^*). \quad (5.18)$$

The last average to compute is over the source directions $\mathbf{\Omega}$.

The average over $\mathbf{\Omega}$ is defined by the final set of integrals in (4.7). We assume that the pulsar distances from Earth, and the distances between the pulsars, are much greater than the wavelength of the GWs. Thus, $L_1 \gg 2\pi/\omega$, $L_2 \gg 2\pi/\omega$, and $|L_1\hat{\mathbf{p}}_1 - L_2\hat{\mathbf{p}}_2| \gg 2\pi/\omega$. Examining (5.18), one can see from (5.6) and (5.11) that $f_{1,j}^* f_{2,j}$ and $f_{1,j}f_{2,j}^*$ each give four terms. Three are rapidly oscillating functions of $\mathbf{\Omega}_j$. When multiplied in (5.18) by the slowly varying antenna pattern functions $F$, they average to zero. The fourth terms are the only ones that survive, and correspond to setting $f_{1,j}^* f_{2,j}$ and $f_{1,j}f_{2,j}^*$ to unity in (5.18). This gives

$$\langle\rho\rangle = \frac{1}{10}\sum_j A_j^2 \langle F_{1,j}F_{2,j}^* + F_{1,j}^* F_{2,j}\rangle_\mathbf{\Omega}. \quad (5.19)$$

This spherical average of the antenna pattern functions was first evaluated in 1983 by Hellings and Downs [18].

We briefly digress to contrast the current calculation of $\langle\rho\rangle$ to the earlier calculation of $\langle s \rangle$ in Sec. IV A. There, following (4.11), we average the product $e^{ab}(\mathbf{\Omega})e^*_{ab}(\mathbf{\Omega})$ over $\mathbf{\Omega}$. Here, those polarization tensors are replaced by antenna pattern functions, which have a more elaborate dependence upon $\mathbf{\Omega}$.

The spherical average in (5.19) is called the HD curve, and is defined by

$$\mu_u(\gamma) \equiv \langle F_{1,j}F_{2,j}^*\rangle_\mathbf{\Omega} = \langle F_{1,j}^* F_{2,j}\rangle_\mathbf{\Omega}. \quad (5.20)$$

Here, $\gamma$ is the angular separation between the two pulsars, so $\cos\gamma = \hat{\mathbf{p}}_1 \cdot \hat{\mathbf{p}}_2$. The function $\mu_u(\gamma)$ is the mean value of the HD correlation for an unpolarized unit-amplitude source. It is computed explicitly in [9] [see Eq. (3.35) and Appendix D] as

$$\mu_u(\gamma) = \frac{1}{4} + \frac{1}{12}\cos\gamma$$
$$+ \frac{1}{2}(1-\cos\gamma)\log\left(\frac{1-\cos\gamma}{2}\right). \quad (5.21)$$

Thus, we obtain the final expression for the mean value, or first moment,

$$\langle\rho\rangle = \frac{1}{5}\left(\sum_j A_j^2\right)\mu_u(\gamma). \quad (5.22)$$

Not surprisingly, this is proportional to the standard HD correlation function.

### B. Second moment and variance of $\rho$

We proceed in analogy with the earlier computation of $\langle s^2 \rangle$ given in Sec. IV B. We first define

$$P_{1,j} \equiv U_j e^{2i\psi_j} F_{1,j} + V_j e^{-2i\psi_j} F_{1,j}^*,$$
$$P_{2,j} \equiv U_j e^{2i\psi_j} F_{2,j} + V_j e^{-2i\psi_j} F_{2,j}^*, \quad (5.23)$$

where the subscript 1 or 2 denotes the pulsar and $j$ labels the source. Starting with (5.15), we can then write the correlation of the two pulsars' redshifts as

$$\rho = \frac{1}{64}\sum_j \left[P_{1,j}P_{2,j}^* f_{1,j}^* f_{2,j} + P_{1,j}^* P_{2,j} f_{1,j} f_{2,j}^*\right]$$
$$+ \frac{1}{64}\sum_{j\neq k}\left[P_{1,j}P_{2,k}^* f_{1,j}^* f_{2,k} e^{i(\phi_j - \phi_k)}\right.$$
$$\left. + P_{1,j}^* P_{2,k} f_{1,j} f_{2,k}^* e^{-i(\phi_j - \phi_k)}\right]. \quad (5.24)$$

The first line is the diagonal terms ($j = k$) and the second line is the off-diagonal ($j \neq k$) ones.

The first step is to average $\rho^2$ over the phases $\phi$. The procedure is the same one used for (4.13). The quantity $\rho^2$ has three terms, but averaging eliminates the cross term. Then, the average of exponential factors (4.14) introduces Kronecker deltas, eliminating one set of double sums. We are left with

$$\langle\rho^2\rangle_\phi = \frac{1}{4096}\left(\sum_j \left[P_{1,j}P_{2,j}^* f_{1,j}^* f_{2,j} + P_{1,j}^* P_{2,j} f_{1,j} f_{2,j}^*\right]\right)^2$$
$$+ \frac{1}{4096}\sum_{j\neq k}\left[P_{1,j}P_{2,k}^* P_{1,k}P_{2,j}^* f_{1,j}^* f_{2,k} f_{1,k}^* f_{2,j} + P_{1,j}^* P_{2,k} P_{1,k}^* P_{2,j} f_{1,j} f_{2,k}^* f_{1,k} f_{2,j}^*\right.$$
$$\left. + P_{1,j}P_{2,k}^* P_{1,j}^* P_{2,k} f_{1,j}^* f_{2,k} f_{1,j} f_{2,k}^* + P_{1,k}P_{2,j}^* P_{1,k}^* P_{2,j} f_{1,k}^* f_{2,j} f_{1,k} f_{2,j}^*\right]. \quad (5.25)$$

Note that within the double sum, the final two terms are identical, and just correspond to swapping $j$ and $k$. If we now explicitly square the single-sum term, we obtain

$$\langle \rho^2 \rangle_\phi = \frac{1}{4096} \sum_j \left[ P_{1,j} P_{2,j}^* f_{1,j}^* f_{2,j} + P_{1,j}^* P_{2,j} f_{1,j} f_{2,j}^* \right]^2$$

$$+ \frac{1}{2048} \sum_{j \neq k} \left[ P_{1,j} P_{2,j}^* P_{1,k} P_{2,k}^* f_{1,j}^* f_{2,j} f_{1,k}^* f_{2,k} + P_{1,j}^* P_{2,j} P_{1,k}^* P_{2,k} f_{1,j} f_{2,j}^* f_{1,k} f_{2,k}^* \right.$$

$$\left. + P_{1,j} P_{1,j}^* P_{2,k} P_{2,k}^* f_{1,j}^* f_{1,j} f_{2,k}^* f_{2,k} + P_{1,j} P_{2,j}^* P_{1,k}^* P_{2,k} f_{1,j}^* f_{2,j} f_{1,k} f_{2,k}^* \right]. \tag{5.26}$$

Here, two of the four off-diagonal terms that come from the square of the single-sum term have doubled the terms in the second line of (5.25). The remaining two off-diagonal terms provide the final term that appears in (5.26). We have also reordered the terms, putting the $j$-factors in front of the $k$-factors. This concludes the average over $\phi$.

We next average over $\psi$, carrying out the second set of integrals in (4.7). Expanding the square in the single sum of (5.26) gives terms proportional to $(P_{1,j} P_{2,j}^*)^2$, $(P_{1,j}^* P_{2,j})^2$ and $P_{1,j} P_{1,j}^* P_{2,j} P_{2,j}^*$. From the definition (5.23), we see that $P_j$ is a linear combination of $e^{2i\psi_j}$ and $e^{-2i\psi_j}$. Thus, only the $\psi$-independent terms survive the average. From the single sum these are

$$\langle (P_{1,j} P_{2,j}^*)^2 \rangle_\psi = U_j^4 (F_{1,j} F_{2,j}^*)^2 + V_j^4 (F_{1,j}^* F_{2,j})^2 + 4U_j^2 V_j^2 |F_{1,j}|^2 |F_{2,j}|^2,$$

$$\langle (P_{1,j}^* P_{2,j})^2 \rangle_\psi = U_j^4 (F_{1,j}^* F_{2,j})^2 + V_j^4 (F_{1,j} F_{2,j}^*)^2 + 4U_j^2 V_j^2 |F_{1,j}|^2 |F_{2,j}|^2,$$

$$2\langle P_{1,j} P_{1,j}^* P_{2,j} P_{2,j}^* \rangle_\psi = 2(U_j^4 + V_j^4 + 2U_j^2 V_j^2) |F_{1,j}|^2 |F_{2,j}|^2 + 2U_j^2 V_j^2 [(F_{1,j} F_{2,j}^*)^2 + (F_{1,j}^* F_{2,j})^2]. \tag{5.27}$$

Since $j \neq k$, the averages over $\psi_j$ and $\psi_k$ may be carried out separately in the double sum of (5.26). In each case, we always have a product of two $P$ terms. In the same way that we computed (5.17), the $e^{\pm 4i\psi}$ terms average to zero. Thus, we obtain

$$\langle \langle \rho^2 \rangle_\phi \rangle_\psi = \frac{1}{4096} \sum_j \left[ \left( U_j^4 (F_{1,j} F_{2,j}^*)^2 + V_j^4 (F_{1,j}^* F_{2,j})^2 + 4U_j^2 V_j^2 |F_{1,j}|^2 |F_{2,j}|^2 \right) (f_{1,j}^* f_{2,j})^2 \right.$$

$$+ \left( U_j^4 (F_{1,j}^* F_{2,j})^2 + V_j^4 (F_{1,j} F_{2,j}^*)^2 + 4U_j^2 V_j^2 |F_{1,j}|^2 |F_{2,j}|^2 \right) (f_{1,j} f_{2,j}^*)^2$$

$$\left. + 2 \left( (U_j^4 + V_j^4 + 2U_j^2 V_j^2) |F_{1,j}|^2 |F_{2,j}|^2 + U_j^2 V_j^2 ((F_{1,j} F_{2,j}^*)^2 + (F_{1,j}^* F_{2,j})^2) \right) |f_{1,j}|^2 |f_{2,j}|^2 \right]$$

$$+ \frac{1}{2048} \sum_{j \neq k} \left[ (U_j^2 F_{1,j} F_{2,j}^* + V_j^2 F_{1,j}^* F_{2,j})(U_k^2 F_{1,k} F_{2,k}^* + V_k^2 F_{1,k}^* F_{2,k}) f_{1,j}^* f_{2,j} f_{1,k}^* f_{2,k} \right.$$

$$+ (U_j^2 F_{1,j}^* F_{2,j} + V_j^2 F_{1,j} F_{2,j}^*)(U_k^2 F_{1,k}^* F_{2,k} + V_k^2 F_{1,k} F_{2,k}^*) f_{1,j} f_{2,j}^* f_{1,k} f_{2,k}^*$$

$$+ (U_j^2 + V_j^2)(U_k^2 + V_k^2) |F_{1,j}|^2 |F_{2,k}|^2 |f_{1,j}|^2 |f_{2,k}|^2$$

$$\left. + (U_j^2 F_{1,j} F_{2,j}^* + V_j^2 F_{1,j}^* F_{2,j})(U_k^2 F_{1,k}^* F_{2,k} + V_k^2 F_{1,k} F_{2,k}^*) f_{1,j}^* f_{2,j} f_{1,k} f_{2,k}^* \right]. \tag{5.28}$$

This concludes the average over $\psi$.

The average over $\iota$ follows immediately from (4.18), since the $\iota$ dependence enters via $U$ and $V$. Averaging (5.28) gives

$$\langle \langle \langle \rho^2 \rangle_\phi \rangle_\psi \rangle_\iota$$

$$= \frac{1}{5040} \sum_j A_j^4 \left[ \left( 35(F_{1,j} F_{2,j}^*)^2 + 35(F_{1,j}^* F_{2,j})^2 + 2|F_{1,j}|^2 |F_{2,j}|^2 \right) \left( (f_{1,j}^* f_{2,j})^2 + (f_{1,j} f_{2,j}^*)^2 \right) \right.$$

$$\left. + \left( (F_{1,j} F_{2,j}^*)^2 + (F_{1,j}^* F_{2,j})^2 + 142|F_{1,j}|^2 |F_{2,j}|^2 \right) |f_{1,j}|^2 |f_{2,j}|^2 \right]$$

$$+ \frac{1}{200} \sum_{j \neq k} A_j^2 A_k^2 \left[ \left( F_{1,j} F_{2,j}^* + F_{1,j}^* F_{2,j} \right) \left( F_{1,k} F_{2,k}^* + F_{1,k}^* F_{2,k} \right) \left( f_{1,j}^* f_{2,j} f_{1,k}^* f_{2,k} + f_{1,j} f_{2,j}^* f_{1,k} f_{2,k}^* + f_{1,j}^* f_{2,j} f_{1,k} f_{2,k}^* \right) \right.$$

$$\left. + 4|F_{1,j}|^2 |F_{2,k}|^2 |f_{1,j}|^2 |f_{2,k}|^2 \right]. \tag{5.29}$$

Some care is needed to average (5.29) over the source directions.

For the final average over $\mathbf{\Omega}$, we make the same assumptions about pulsar and interpulsar distances as given before (5.19). Hence, the situation in (5.29) is similar. The antenna pattern functions $F$ vary slowly with $\mathbf{\Omega}$, whereas the phase difference terms $f$ vary quickly. Thus, the same reasoning previously used to obtain (5.19) can also be used here. It implies that the average of terms which do not contain $|f|^2$ are obtained by setting $f \to 1$. In contrast, the average of terms that contain $|f|^2$ are obtained by setting $|f|^2 \to 1 + \chi^2$. See Eqs. (3.34) and (3.41) of [9] for a more detailed account. So, in the end, all that is needed is averages of the antenna pattern functions $F$ over source direction.

The necessary antenna pattern averages can be obtained from [9], where they are given in terms of plus and cross polarizations. We use (5.10) to obtain the complex polarization equivalents. These first and second moments are

$$\langle F_{1,j} F_{2,j}^* \rangle_{\mathbf{\Omega}_j} = \mu_\mathrm{u}(\gamma), \tag{5.30}$$

$$\mathfrak{R}\langle (F_{1,j} F_{2,j}^*)^2 \rangle_{\mathbf{\Omega}_j} = \mu_\mathrm{u}^2(\gamma) + \sigma_\mathrm{u}^2(\gamma) - \sigma_\mathrm{p}^2(\gamma), \tag{5.31}$$

$$\langle |F_{1,j}|^2 |F_{2,j}|^2 \rangle_{\mathbf{\Omega}_j} = \mu_\mathrm{u}^2(\gamma) + \sigma_\mathrm{u}^2(\gamma) + \sigma_\mathrm{p}^2(\gamma). \tag{5.32}$$

As before, $\gamma$ is the angle between the pulsar directions $\hat{\mathbf{p}}_1$ and $\hat{\mathbf{p}}_2$. Only the real part is given in (5.31), because (5.29) does not contain the imaginary part. The HD curve function $\mu_\mathrm{u}(\gamma)$ was already given in (5.20), and the variance functions $\sigma_\mathrm{u}^2(\gamma)$ and $\sigma_\mathrm{p}^2(\gamma)$ are given in Eqs. (D37) and (E8) of [9].

We use relations (5.30)–(5.32) to carry out the final average of (5.29) over source directions $\mathbf{\Omega}$. In the single-sum term, both (5.31) and (5.32) are required. In the double-sum term, since $j \neq k$, only (5.30) is needed. Note that the average of $|F_{1,j}|^2$ (or $|F_{2,k}|^2$) is obtained from (5.30) by setting $\gamma = 0$. We obtain

$$\begin{aligned}\langle \rho^2 \rangle &= \frac{1}{1260}\left[36\mu_\mathrm{u}^2(\gamma) + 36\sigma_\mathrm{u}^2(\gamma) - 34\sigma_\mathrm{p}^2(\gamma) + (1+\chi^2)^2\left(36\mu_\mathrm{u}^2(\gamma) + 36\sigma_\mathrm{u}^2(\gamma) + 35\sigma_\mathrm{p}^2(\gamma)\right)\right]\sum_j A_j^4 \\ &\quad + \frac{1}{50}\left[3\mu_\mathrm{u}^2(\gamma) + (1+\chi^2)^2 \mu_\mathrm{u}^2(0)\right]\sum_{j\neq k} A_j^2 A_k^2 \\ &= \frac{1}{6300}\left[-198\mu_\mathrm{u}^2(\gamma) + 180\sigma_\mathrm{u}^2(\gamma) - 170\sigma_\mathrm{p}^2(\gamma) + (1+\chi^2)^2\left(180\mu_\mathrm{u}^2(\gamma) + 180\sigma_\mathrm{u}^2(\gamma) + 175\sigma_\mathrm{p}^2(\gamma) - 126\mu_\mathrm{u}^2(0)\right)\right]\sum_j A_j^4 \\ &\quad + \frac{1}{50}\left[3\mu_\mathrm{u}^2(\gamma) + (1+\chi^2)^2 \mu_\mathrm{u}^2(0)\right]\left(\sum_j A_j^2\right)^2,\end{aligned} \tag{5.33}$$

where the second equality follows from (4.22). This concludes the calculation of the second moment of the correlation $\rho$.

The variance $\sigma^2 = \langle \rho^2 \rangle - \langle \rho \rangle^2$ is obtained from (5.33), and from the first moment (5.22). Thus, for the polarized ensemble defined by (2.4), the mean and variance of the correlation between the pulsars 1 and 2 are

$$\begin{aligned}\langle \rho \rangle_\mathrm{pol} &= \frac{1}{5}\left(\sum_j A_j^2\right)\mu_\mathrm{u}(\gamma), \\ \sigma_\mathrm{pol}^2 &= \frac{1}{3150}\left[261\mu_\mathrm{u}^2(\gamma) + 450\sigma_\mathrm{u}^2(\gamma) + 265\sigma_\mathrm{p}^2(\gamma) - 252\mu_\mathrm{u}^2(0)\right]\sum_j A_j^4 + \frac{1}{50}\left[\mu_\mathrm{u}^2(\gamma) + 4\mu_\mathrm{u}^2(0)\right]\left(\sum_j A_j^2\right)^2,\end{aligned} \tag{5.34}$$

where the sums are over the $N$ sources and we set $\chi = 1$, which gives the correct physical description of a PTA. To see the effects of discarding the pulsar terms, set $\chi = 0$, as discussed after (5.1).

### C. Comparison of polarized and unpolarized ensembles

We now compare these results for the polarized ensemble to the corresponding results for the unpolarized ensemble. The mean and variance of the correlation $\rho$ for the unpolarized ensemble (2.1) are given in Eqs. (3.36) and (3.48) of [9]. Setting $\chi = 1$ gives

$$\begin{aligned}\langle \rho \rangle_\mathrm{unpol} &= \frac{1}{2}\left(\sum_j \mathcal{A}_j^2\right)\mu_\mathrm{u}(\gamma), \\ \sigma_\mathrm{unpol}^2 &= \frac{1}{8}\left(2\mu_\mathrm{u}^2(\gamma) + 5\sigma_\mathrm{u}^2(\gamma) + 3\sigma_\mathrm{p}^2(\gamma) - 4\mu_\mathrm{u}^2(0)\right)\sum_j \mathcal{A}_j^4 + \frac{1}{8}\left(\mu_\mathrm{u}^2(\gamma) + 4\mu_\mathrm{u}^2(0)\right)\left(\sum_j \mathcal{A}_j^2\right)^2.\end{aligned} \tag{5.35}$$

To obtain this, we have used the relation

$$\sigma_c^2(\gamma) = \sigma_p^2(\gamma) + \sigma_u^2(\gamma) + \mu_u^2(\gamma), \quad (5.36)$$

which is given in Eq. (F4) of [9].

As in Sec. IV C, to compare the two ensembles on "equal footing" we assume that the amplitudes of the individual GW sources are related by (4.25). With this, we find

$$\langle \rho \rangle_\text{pol} = \langle \rho \rangle_\text{unpol},$$
$$\sigma_\text{pol}^2(\gamma) = \sigma_\text{unpol}^2(\gamma)$$
$$+ \left( \frac{3}{70} \mu_u^2(\gamma) + \frac{3}{70} \sigma_u^2(\gamma) + \frac{38}{1575} \sigma_p^2(\gamma) \right) \sum_j A_j^4. \quad (5.37)$$

The additional terms that appear on the rhs are *positive*. This follows from general principles: introducing additional degrees of freedom (here $\iota_j$ and $\psi_j$) increases the variance.

There are two limits of interest. First, for a single polarized source, this result reduces to that given in Eq. (B11) and footnote 43 of [9]. Second, for a very large number of polarized sources, the $(\sum_j A_j^2)^2$ terms dominate the $\sum_j A_j^4$ terms, hence in this limit $\sigma_\text{pol}^2/\sigma_\text{unpol}^2 \to 1$.

## VI. COSMIC VARIANCE

The variance $\sigma_\text{pol}^2$ in (5.34) quantifies the *total* range of variation in the correlation between two pulsars. This variance is the sum of *pulsar variance* and *cosmic variance*.

For a given set of GW sources, different pairs of pulsars with the same angular separation will exhibit different correlations. This variation, associated with picking different pulsar pairs separated by the same angle, is the pulsar variance. In contrast, the cosmic variance arises because any given set of GW sources differs from the mean.

The pulsar variance can be removed from the total variance by averaging the correlation $\rho$ over all pairs of pulsars separated by the same angle $\gamma$. This replaces the correlation $\rho$ with a *pulsar-averaged* correlation $\Gamma(\gamma)$, where $\Gamma(\gamma)$ is the average of $\rho$ over all pairs of pulsars separated by the same angle $\gamma$. (By definition) the variance of $\Gamma(\gamma)$ is the cosmic variance [9].

We thus define the pulsar-averaged correlation as

$$\Gamma(\gamma) \equiv \langle \rho \rangle_\text{p}. \quad (6.1)$$

Here, the subscript "p" denotes an average over all pulsar pairs $\hat{\mathbf{p}}_1$ and $\hat{\mathbf{p}}_2$ separated by $\gamma$, where $\hat{\mathbf{p}}_1$ and $\hat{\mathbf{p}}_2$ are uniformly distributed over the sphere. The only terms in the correlation $\rho$ (5.15) which depend upon the pulsar positions are the antenna pattern functions $F_j$ (5.10) and the phase difference terms $f_j$ (5.11).

The products of $f$ reduce to unity when computing the pulsar average. This is because of our assumption that the pulsars 1 and 2 are separated by many gravitational wavelengths. Each $f$ includes a rapidly oscillating complex exponential function of the pulsar direction $\hat{\mathbf{p}}$ (5.11). The product of this function with the slowly varying $F$ averages to zero. Thus, provided that pulsars 1 and 2 are separated by many gravitational wavelengths, only the Earth terms survive. Since the Earth terms are unity, this implies

$$\langle F_{1,j}^* F_{2,k} f_{1,j} f_{2,k}^* \rangle_\text{p} = \langle F_{1,j}^* F_{2,k} \rangle_\text{p}. \quad (6.2)$$

The same reasoning applies to products where fewer or more of the terms are complex conjugated.

The only terms left to average to obtain $\Gamma(\gamma)$ from (5.15) are products of antenna pattern functions. Those averages are calculated in Appendix G of [9]. The results are

$$\langle F_{1,j} F_{2,k}^* \rangle_\text{p} = \langle F_{1,j}^* F_{2,k} \rangle_\text{p}^* = \mu(\gamma, \beta_{jk}) e^{2i\chi_{jk}},$$
$$\langle F_{1,j} F_{2,k} \rangle_\text{p}^* = \langle F_{1,j}^* F_{2,k}^* \rangle_\text{p} = \mu(\pi-\gamma, \pi-\beta_{jk}) e^{2i\bar{\chi}_{jk}}. \quad (6.3)$$

Here, $\gamma$ is the angular separation between the directions to the two pulsars, $\beta_{jk} = \beta_{kj}$ is the angular separation between the direction to the two sources $j$ and $k$, so $\cos\beta_{jk} = \mathbf{\Omega}_j \cdot \mathbf{\Omega}_k$ and $\chi_{jk}$ and $\bar{\chi}_{jk}$ are real angles satisfying $\chi_{jk} = -\chi_{kj}$ and $\bar{\chi}_{jk} = -\bar{\chi}_{kj}$. The two-point function $\mu(\gamma, \beta)$ is real, and given in Eq. (4.2) of [9]. It is a generalization of the HD curve $\mu_u(\gamma)$; the two functions coincide in the case where $\beta_{jk}$ vanishes: $\mu(\gamma, 0) = \mu_u(\gamma)$. [Note: the phases $\chi_{jk}$ and $\bar{\chi}_{jk}$ are gauge artifacts that were inadvertently set to zero throughout [9]. We have also omitted them in what follows. It is easy to demonstrate that after computing the average $\langle \rangle_\phi$, both $\chi_{jk}$ and $\bar{\chi}_{jk}$ cancel out of the first moment (6.5) and the second moment (6.7). See [19] for details.]

We now calculate the pulsar-averaged correlation $\Gamma(\gamma)$. Using (6.2) and (6.3) to average (5.15) over all pulsar pairs gives

$$\Gamma(\gamma) = \frac{1}{64} \sum_{j,k} \left[ \left( (U_j V_k e^{2i(\psi_j+\psi_k)} + V_j U_k e^{-2i(\psi_j+\psi_k)}) e^{i(\phi_j-\phi_k)} \right. \right.$$
$$\left. + (V_j U_k e^{2i(\psi_j+\psi_k)} + U_j V_k e^{-2i(\psi_j+\psi_k)}) e^{-i(\phi_j-\phi_k)} \right) \mu(\pi-\gamma, \pi-\beta_{jk})$$
$$+ \left( (U_j U_k e^{2i(\psi_j-\psi_k)} + V_j V_k e^{-2i(\psi_j-\psi_k)}) e^{i(\phi_j-\phi_k)} \right.$$
$$\left. \left. + (V_j V_k e^{2i(\psi_j-\psi_k)} + U_j U_k e^{-2i(\psi_j-\psi_k)}) e^{-i(\phi_j-\phi_k)} \right) \mu(\gamma, \beta_{jk}) \right]. \quad (6.4)$$

We now compute the mean and variance of $\Gamma$.

### A. First moment of $\Gamma$

Calculating the ensemble average of $\Gamma(\gamma)$ is very similar to the calculation of $\langle \rho \rangle$ in Sec. V A. We proceed by averaging (6.4) over the source ensemble as described in (4.7).

We start by calculating the average over the GW source phases $\phi_j$, evaluating the first set of integrals in (4.7). This gives

$$\langle \Gamma(\gamma) \rangle_\phi = \frac{1}{32} \sum_j \Big[ (U_j^2 + V_j^2)\mu(\gamma, \beta_{jj})$$
$$+ U_j V_j (e^{4i\psi_j} + e^{-4i\psi_j})\mu(\pi - \gamma, \pi - \beta_{jj}) \Big]$$
$$= \frac{1}{32} \sum_j (U_j^2 + V_j^2)\mu_u(\gamma), \quad (6.5)$$

where only the diagonal terms ($j = k$) survived the average. Note that since $\beta_{jj} = 0$, the two-point function $\mu(\gamma, \beta_{jj}) = \mu_u(\gamma)$ and $\mu(\pi - \gamma, \pi - \beta_{jj}) = \mu(\pi - \gamma, \pi) = 0$. (For the final equality, see for example Fig. 12 in Appendix G of [9].) As expected, the result is proportional to the HD curve $\mu_u(\gamma)$.

Since the final line of (6.5) is independent of $\psi_j$, the only averages remaining in (4.7) are over $\iota_j$ and $\mathbf{\Omega}_j$. The quantities that appear in the final line of (6.5) are independent of $\mathbf{\Omega}_j$, because $U_j$ and $V_j$ only depend upon $\iota_j$. Using (4.18) for the averaging gives:

$$\langle \Gamma(\gamma) \rangle = \frac{1}{5}\left( \sum_j A_j^2 \right) \mu_u(\gamma). \quad (6.6)$$

Thus, the expectation value of the pulsar-averaged correlation $\Gamma(\gamma)$ in (6.6) is the same as the expectation value of $\rho$ in (5.22). This is not surprising: *on average* an observer would always obtain a correlation curve proportional to the HD function.

### B. Second moment and variance of $\Gamma$

We now compute the second moment and the variance of $\Gamma(\gamma)$. The calculation is very similar to the one carried out in Sec. V B for $\rho$, and corresponds again to evaluating the average given by (4.7).

We begin by computing $\Gamma^2(\gamma)$. As we did for $\rho$ in (5.24), the sum in (6.4) is first divided into diagonal terms ($j = k$) and the off-diagonal ones ($j \neq k$). $\Gamma^2(\gamma)$ then consists of three terms: (i) the square of the diagonal sum, (ii) the square of the off-diagonal sum, and (iii) the cross-product terms.

We then average $\Gamma^2(\gamma)$ over the phases $\phi_j$. Two of the three terms are easy to average: (i) does not change because it is independent of $\phi_j$, and (iii) vanishes. The square of the off-diagonal sum (ii) contains the product of two double sums: $\sum_{j \neq k} \sum_{\ell \neq m}$. This double sum has 16 terms, each one containing a complex exponential $e^{\pm i((\phi_j - \phi_k) \pm (\phi_\ell - \phi_m))}$. Using (4.14), (ii) reduces to a sum over $j \neq k$, and the 16 terms reduce to 4. We obtain

$$\langle \Gamma^2(\gamma) \rangle_\phi = \frac{1}{1024}\left( \sum_j (U_j^2 + V_j^2)\mu_u(\gamma) \right)^2$$
$$+ \frac{1}{1024} \sum_{j \neq k} \Big[ \left( U_j V_k e^{2i(\psi_j + \psi_k)} + V_j U_k e^{-2i(\psi_j + \psi_k)} \right)\left( V_j U_k e^{2i(\psi_j + \psi_k)} + U_j V_k e^{-2i(\psi_j + \psi_k)} \right)\mu^2(\pi - \gamma, \pi - \beta_{jk})$$
$$+ \left( U_j V_k e^{2i(\psi_j + \psi_k)} + V_j U_k e^{-2i(\psi_j + \psi_k)} \right)\left( V_j V_k e^{2i(\psi_j - \psi_k)} + U_j U_k e^{-2i(\psi_j - \psi_k)} \right)\mu(\pi - \gamma, \pi - \beta_{jk})\mu(\gamma, \beta_{jk})$$
$$+ \left( U_j U_k e^{2i(\psi_j - \psi_k)} + V_j V_k e^{-2i(\psi_j - \psi_k)} \right)\left( V_j U_k e^{2i(\psi_j + \psi_k)} + U_j V_k e^{-2i(\psi_j + \psi_k)} \right)\mu(\pi - \gamma, \pi - \beta_{jk})\mu(\gamma, \beta_{jk})$$
$$+ \left( U_j U_k e^{2i(\psi_j - \psi_k)} + V_j V_k e^{-2i(\psi_j - \psi_k)} \right)\left( V_j V_k e^{2i(\psi_j - \psi_k)} + U_j U_k e^{-2i(\psi_j - \psi_k)} \right)\mu^2(\gamma, \beta_{jk}) \Big]. \quad (6.7)$$

This concludes the average over $\phi_j$.

The next step is the average over $\psi_j$. Carrying out the second set of integrals in (4.7), the second and third terms of the double sum of (6.7) vanish. We obtain

$$\langle \langle \Gamma^2(\gamma) \rangle_\phi \rangle_\psi = \frac{1}{1024}\Bigg\{ \sum_{j \neq k} \Big[ (U_j^2 + V_j^2)(U_k^2 + V_k^2)\mu_u^2(\gamma) + 2U_j^2 V_k^2 \mu^2(\pi - \gamma, \pi - \beta_{jk}) + (U_j^2 U_k^2 + V_j^2 V_k^2)\mu^2(\gamma, \beta_{jk}) \Big]$$
$$+ \sum_j (U_j^2 + V_j^2)^2 \mu_u^2(\gamma) \Bigg\}, \quad (6.8)$$

where we also expanded the square of the single sum in the first line of (6.7) and exploited the symmetry of the double sum under the interchange of $j$ and $k$.

We now average over $\iota_j$. The only terms depending upon $\iota_j$ are the functions $U_j$ and $V_j$. Using (4.18), we obtain

$$\langle\langle\langle\Gamma^2(\gamma)\rangle_\phi\rangle_\psi\rangle_\iota = \frac{71}{1260}\mu_u^2(\gamma)\sum_j A_j^4 + \frac{1}{50}\sum_{j\neq k}\Big(2\mu_u^2(\gamma) + \mu^2(\pi-\gamma,\pi-\beta_{jk}) + \mu^2(\gamma,\beta_{jk})\Big)A_j^2 A_k^2. \tag{6.9}$$

This concludes the average over $\iota_j$.

The remaining step to evaluate the second moment is to average over the source directions $\mathbf{\Omega}_j$. The dependence on $\mathbf{\Omega}_j$ is via $\beta_{jk}$, where $\cos(\beta_{jk}) = \mathbf{\Omega}_j \cdot \mathbf{\Omega}_k$. The spherical averages over $\mathbf{\Omega}_j$ and $\mathbf{\Omega}_k$ are equivalent to evaluating the average over $\cos\beta_{jk}$. To visualize this, assume that the $j$th source is aligned with the $\hat{z}$ axis. Thus, $\beta_{jk} = \beta_{\hat{z}k} \equiv \theta$, where $\theta$ is the traditional polar angle in spherical polar coordinates. Then, for any function $g(\beta_{jk})$,

$$\int\frac{d\mathbf{\Omega}_j}{4\pi}\int\frac{d\mathbf{\Omega}_k}{4\pi}g(\beta_{jk}) = \int\frac{d\mathbf{\Omega}_k}{4\pi}g(\beta_{\hat{z}k}) = \frac{1}{4\pi}\int_0^{2\pi}d\phi\int_{-1}^1 d(\cos\theta)g(\theta) = \frac{1}{2}\int_{-1}^1 g(\theta)d(\cos\theta). \tag{6.10}$$

We can now use (6.10) to average (6.9) over the source directions.

The only terms depending upon $\beta_{jk}$ are $\mu^2(\pi-\gamma,\pi-\beta_{jk})$ and $\mu^2(\gamma,\beta_{jk})$. Those averages are discussed in detail and computed in Appendix G of [9]. Starting with the second line of Eq. (G12) of [9] and making use of (6.10) we have

$$\int\frac{d\mathbf{\Omega}_j}{4\pi}\int\frac{d\mathbf{\Omega}_k}{4\pi}\Big(\mu^2(\gamma,\beta_{jk}) + \mu^2(\pi-\gamma,\pi-\beta_{jk})\Big) = \frac{1}{2}\int_{-1}^1\Big(\mu^2(\gamma,\beta) + \mu^2(\pi-\gamma,\pi-\beta)\Big)d(\cos\beta) = 2\tilde{\mu}^2(\gamma), \tag{6.11}$$

where the complete expression for $\tilde{\mu}^2(\gamma)$ is given in Eq. (G11) of [9]. Using (6.11) to average (6.9) over the source directions then gives

$$\langle\Gamma^2(\gamma)\rangle = \frac{71}{1260}\mu_u^2(\gamma)\sum_j A_j^4 + \frac{1}{25}\Big(\mu_u^2(\gamma) + \tilde{\mu}^2(\gamma)\Big)\sum_{j\neq k}A_j^2 A_k^2$$
$$= \Big(\frac{103}{6300}\mu_u^2(\gamma) - \frac{1}{25}\tilde{\mu}^2(\gamma)\Big)\sum_j A_j^4 + \frac{1}{25}\Big(\mu_u^2(\gamma) + \tilde{\mu}^2(\gamma)\Big)\Big(\sum_j A_j^2\Big)^2, \tag{6.12}$$

where we used (4.22) to obtain the second line. This completes the calculation of the second moment of $\Gamma(\gamma)$.

We can finally write the variance of the pulsar-averaged correlation $\Gamma(\gamma)$. From (6.6) and (6.12), we obtain

$$\sigma_{\text{cosmic}\atop\text{pol}}^2 \equiv \langle\Gamma^2(\gamma)\rangle - \langle\Gamma(\gamma)\rangle^2$$
$$= \Big(\frac{103}{6300}\mu_u^2(\gamma) - \frac{1}{25}\tilde{\mu}^2(\gamma)\Big)\sum_j A_j^4$$
$$+ \frac{1}{25}\tilde{\mu}^2(\gamma)\Big(\sum_j A_j^2\Big)^2, \tag{6.13}$$

which corresponds to the *cosmic variance* of the correlation $\rho$ (5.15) for the polarized ensemble defined by (2.4).

### C. Comparison of polarized and unpolarized ensembles

We now compare the cosmic variance (6.13) of the polarized ensemble (2.4) with that obtained for the unpolarized ensemble (2.1).

The cosmic variance for the unpolarized ensemble (2.1) is given by Eq. (4.8) of [9] as

$$\sigma_{\text{cosmic}\atop\text{unpol}}^2 = \frac{1}{4}\tilde{\mu}^2(\gamma)\Big(\Big(\sum_j \mathcal{A}_j^2\Big)^2 - \sum_j \mathcal{A}_j^4\Big), \tag{6.14}$$

where $\tilde{\mu}^2(\gamma)$ is the same function as the one introduced in (6.11).

As in Secs. IV C and V C, to compare the two ensembles on "equal footing," we assume that the amplitudes of the individual GW sources are related by (4.25). Thus, we obtain

$$\sigma_{\text{cosmic}\atop\text{pol}}^2 = \sigma_{\text{cosmic}\atop\text{unpol}}^2 + \frac{103}{6300}\mu_u^2(\gamma)\sum_j A_j^4. \tag{6.15}$$

The additional term of the polarized ensemble variance is *positive*. Thus, the polarized ensemble has a larger cosmic variance than the unpolarized one. This is to be expected: see the discussion in Sec. V C. We note that the additional degrees of freedom have increased the variance "in quadrature" by a term proportional to the square of the mean. There is probably a simple physical explanation for this.

In the many-source limit $N \to \infty$, the ratio of the two cosmic variances approaches unity, $\sigma^2_{\text{cos,pol}}/\sigma^2_{\text{cos,unpol}} \to 1$. The same behavior was also found for the total variance in Sec. V C. This follows immediately from (6.15), because the sum $\sum_j A_j^2$ diverges as the number of sources grows, whereas the sum $\sum_j A_j^4$ converges to $A^4 \zeta(4/3)$. Thus, in the many-source limit, the two ensembles have the same behavior.

## VII. COMPARISON WITH A GAUSSIAN ENSEMBLE

The frequency-domain Gaussian ensemble (introduced in [6,7]) is often used to describe PTA sources. For example, the ENTERPRISE software package [5] used to analyze PTA data assumes that the sources are described by a Gaussian ensemble.

Here, we compare our results for an ensemble of $N$ discrete sources to the corresponding results for a Gaussian ensemble. Specifically, we compare the HD-correlation mean, total variance and cosmic variance of our ensemble of polarized sources to those of the Gaussian ensemble. These are derived in Appendix C of [9].

In the weak-signal limit, the gravitational wave amplitude can always be written as a plane wave expansion

$$h_{ab}(t, \mathbf{x}) = \sum_A \int df \int d\Omega \, e^{2\pi i f(t - \mathbf{\Omega} \cdot \mathbf{x})} h_A(f, \mathbf{\Omega}) e^A_{ab}(\mathbf{\Omega}). \tag{7.1}$$

Here, $A$ denotes the polarization $(+, \times)$, $f \in \mathfrak{R}$ denotes the GW frequency, $\mathbf{\Omega}$ denotes a unit vector in the GW propagation direction, and $e^A_{ab}(\mathbf{\Omega})$ is the polarization tensor defined in Eqs. (D6) and (D7) of [9]. The complex function $h_A(f, \mathbf{\Omega})$ is the GW Fourier amplitude, and differs from one realization to the next. For any realization, the GW amplitude $h_{ab}(t, \mathbf{x})$ is real, implying that $h_A(f, \mathbf{\Omega}) = h_A^*(-f, \mathbf{\Omega})$, where $*$ denotes the complex conjugate.

Under the assumption that the GWs are generated by some central-limit-theorem process, we describe their statistics by treating the functions $h_A(f, \mathbf{\Omega})$ as Gaussian random variables. In this section, we use angle brackets $\langle Q \rangle$ to denote the average of $Q$ in this Gaussian ensemble. Note that this is different than earlier in this paper, where the ensemble average was defined by (4.7). In this section, the ensemble average refers to a Gaussian ensemble of functions $h_A(f, \mathbf{\Omega})$ (see Eq. (C2) of [9]).

The Gaussian ensemble is completely defined by:
(a) The first moment: the ensemble average of the strain vanishes:

$$\langle h_A(f, \mathbf{\Omega}) \rangle = 0. \tag{7.2}$$

(b) The second moment:

$$\langle h_A(f, \mathbf{\Omega}) h_{A'}^*(f', \mathbf{\Omega}') \rangle = \delta_{AA'} \delta^2(\mathbf{\Omega}, \mathbf{\Omega}') \delta(f - f') H(f), \tag{7.3}$$

where $H(f) = H(-f)$ is a real function that defines the measure of the squared amplitude of perturbations at GW frequency $f$ (see Eqs. (C5) and (C7) of [9]). The delta function in frequency implies that the ensemble is second-order stationary in time, while the delta function on the sphere ensures second-order stationarity in space. Lastly, the delta function in polarization implies that the two polarizations are identical, but uncorrelated (the ensemble is unpolarized).

(c) Isserlis' theorem [20]: the higher moments [expected value of the product of $h_A(f, \mathbf{\Omega}) h_{A'}(f', \mathbf{\Omega}') \cdots h_{A''}(f'', \mathbf{\Omega}'')$] are given by sums of expected values of all possible terms containing one or two $h$'s. For example

$$\begin{aligned}
&\langle h_A(f, \mathbf{\Omega}) h_{A'}(f', \mathbf{\Omega}') h_{A''}(f'', \mathbf{\Omega}'') \rangle \\
&= \langle h_A(f, \mathbf{\Omega}) \rangle \langle h_{A'}(f', \mathbf{\Omega}') h_{A''}(f'', \mathbf{\Omega}'') \rangle + \\
&\quad \langle h_A(f, \mathbf{\Omega}) h_{A'}(f', \mathbf{\Omega}') \rangle \langle h_{A''}(f'', \mathbf{\Omega}'') \rangle + \\
&\quad \langle h_{A'}(f', \mathbf{\Omega}') \rangle \langle h_A(f, \mathbf{\Omega}) h_{A''}(f'', \mathbf{\Omega}'') \rangle \\
&= 0, \tag{7.4}
\end{aligned}$$

which vanishes as direct consequence of condition (a), and

$$\begin{aligned}
&\langle h_A(f, \mathbf{\Omega}) h_{A'}(f', \mathbf{\Omega}') h_{A''}(f'', \mathbf{\Omega}'') h_{A'''}(f''', \mathbf{\Omega}''') \rangle \\
&= \langle h_A(f, \mathbf{\Omega}) h_{A'}(f', \mathbf{\Omega}') \rangle \langle h_{A''}(f'', \mathbf{\Omega}'') h_{A'''}(f''', \mathbf{\Omega}''') \rangle + \\
&\quad \langle h_A(f, \mathbf{\Omega}) h_{A''}(f'', \mathbf{\Omega}'') \rangle \langle h_{A'}(f', \mathbf{\Omega}') h_{A'''}(f''', \mathbf{\Omega}''') \rangle + \\
&\quad \langle h_A(f, \mathbf{\Omega}) h_{A'''}(f''', \mathbf{\Omega}''') \rangle \langle h_{A'}(f', \mathbf{\Omega}') h_{A''}(f'', \mathbf{\Omega}'') \rangle.
\end{aligned} \tag{7.5}$$

The ensemble average of the products of more than four $h$'s follows the same pattern, reducing to sums of the expected values of all possible terms containing one or two $h$'s.

Following the same prescription as the ones reported in Sec. V and VI, Appendix C of [9] computes the total variance and cosmic variance for a Gaussian ensemble of GW sources. We compare the results shown in that appendix with the ones estimated for our polarized discrete ensemble of sources.

### A. Correlation mean and total variance: Comparison of polarized and Gaussian ensembles

The mean and total variance of the HD correlation for a Gaussian ensemble (denoted with subscript "G") are

$$\langle \rho \rangle_G = h^2 \mu_u(\gamma),$$
$$\sigma_G^2 = \hbar^4 (\mu_u^2(\gamma) + 4\mu_u^2(0)), \quad (7.6)$$

which correspond to Eqs. (C18) and (C28) of [9]. The scale factors $h$ and $\hbar$ are measures of the strain, expressed as integrals of the spectral function $H(f)$. From Eqs. (C19) and (C26) of [9]:

$$h^2 \equiv 4\pi \int H(f) df,$$
$$\hbar^4 \equiv (4\pi)^2 \int df \int df' \operatorname{sinc}^2(\pi(f-f')T) H(f) H(f'), \quad (7.7)$$

where $T$ is the total observation time.

We now compare the mean correlation and variance estimated for the Gaussian ensemble (7.6) with the results for our polarized ensemble of discrete sources (5.34). To compare the two ensembles "on equal footing," we set the normalizations (GW-amplitude scale factors) to:

$$h^2 \equiv \frac{1}{5} \sum_j A_j^2,$$
$$\hbar^4 \equiv \frac{1}{50} \left( \sum_j A_j^2 \right)^2, \quad (7.8)$$

which implies that $\hbar^4 = h^4/2$.

With this normalization choice (7.8), the expected mean correlation and its variance for the ensemble of polarized sources (5.34) converge to the Gaussian ones (7.6) in the limit of an infinite number of sources. A similar convergence was found in [9] for an ensemble of unpolarized GW sources.

### B. Cosmic variance: Comparison of polarized and Gaussian ensembles

We now compare the cosmic variance for a polarized and a Gaussian ensemble of GW sources.

For a polarized ensemble, the expected value of the pulsar-averaged correlation $\Gamma$ and its variance are given in (6.6) and (6.13). For a Gaussian ensemble of sources, the corresponding quantities are given in Appendix C of [9]:

$$\langle \Gamma(\gamma) \rangle_G = h^2 \mu_u(\gamma),$$
$$\sigma_{G,\text{cosmic}}^2 = 2\hbar^4 \tilde{\mu}^2(\gamma), \quad (7.9)$$

from Eqs. (C42) and (C45) of that reference.

In the limit of an infinite number of sources and with the normalization relationship (7.8), the cosmic variance of the ensemble of polarized sources (6.13) converges to that of the Gaussian ensemble of sources (7.9). A similar convergence was found in [9] for an ensemble of unpolarized GW sources.

## VIII. CONCLUSION

The most common model of the gravitational wave stochastic background is the Gaussian ensemble. However, previous work [9] shows that this does not have the same statistical properties as the gravitational wave background produced by a finite set of discrete sources. To study the discrete-source case, that work constructs an ensemble which is simple, but unrealistic. It is unrealistic because the discrete sources correspond to circular-orbit binary systems, viewed face-on. These produce circularly polarized gravitational waves (2.1). Physically, we expect the orbits to have all possible orientations ranging from face-on to face-off, producing elliptically polarized gravitational waves.

This paper adopts a more realistic model. We employ an ensemble of discrete sources (2.4) described in Sec. II, for which the orbital angular momentum directions are uniformly distributed over the two-sphere of directions. This is a more realistic gravitational wave ensemble, because it corresponds to a set of circular-orbit binary systems with randomly distributed orbital planes. For example, a set of supermassive black-hole binary sources whose circular orbits have inclination angles $\iota$ which are random variables with $\cos \iota$ distributed uniformly between $-1$ and $1$. The previous model assumes $\cos \iota = 1$ (face-on).

To construct the ensemble, in Sec. III we use a set of $N$ distant gravitational waves sources, labeled by an integer $j = 1, \ldots, N$. Their amplitudes fall off with distance as would be expected for identical (intrinsic amplitude) sources, uniformly distributed inside a sphere centered at Earth. Each source has its own orbital phase $\phi_j$, sky direction $-\mathbf{\Omega}_j$, and orbital orientation parameters: $\cos \iota_j$ for inclination and $\psi_j$ for elliptical axis direction. The ensemble is defined by treating these $5N$ parameters (note: $\mathbf{\Omega}$ counts as two) as independent random variables; the ensemble average (4.7) is a product of integrals over them. The discrete ensemble of [9] has only $3N$ independent random variables, because $\iota$ and $\psi$ have fixed values $\iota = \psi = 0$.

The mean and variance of the HD correlation are complicated to calculate. So, before evaluating these, Sec. IV examines a simpler quantity $s$: the time-averaged squared strain. The mean and variance of $s$ are computed, and then compared for the polarized and unpolarized ensembles in Sec. IV C. If the parameters of the two ensembles are set to give the same mean value of $s$, then the variance of $s$ is larger for the polarized ensemble. This variance is "cosmic variance": it describes how $s$ varies between different realizations of the universe.

In Sec. V, using identical techniques, we calculate the two quantities most relevant for pulsar timing arrays. These are the mean and variance of the HD correlation $\rho(\gamma)$, as a function of the angle $\gamma$ between the lines of sight to two pulsars. Later, in Sec. VI, we compute the cosmic variance by first averaging the correlation over many pulsar pairs separated by angle $\gamma$. As with $s$, if the parameters of the ensembles are picked so that the mean values of $\rho(\gamma)$ agree, then the polarized ensemble has larger variance and larger cosmic variance than the unpolarized ensemble. This is not surprising. The varying orbital inclinations and polarizations add additional degrees of freedom, which increase the variance between different realizations drawn from the polarized ensemble.

We also examine the means and variances in the limit of an infinite density of GW sources with the time-averaged squared strain $s$ at Earth held fixed. In all cases, these approach the corresponding quantities for the Gaussian ensemble. This raises an interesting question for future work: is this true for *all* observable quantities? Can we quantify how close the ensembles of discrete sources are to the Gaussian ensemble, and prove that in some limit they approach the Gaussian ensemble?

In the future, galaxy surveys will provide sky direction, distance, mass and frequency information for likely GW sources. The methods employed here can be used to quantify how closely the reconstruction of the HD correlation should match the (mean value prediction of the) HD curve. This work could also be extended to take into account the probability distributions of chirp mass and frequency for the sources, as well as the specific pulsar locations on the sky, as in [16].